\documentclass[aps,prd,onecolumn,11pt,a4paper,tightenlines,eqsecnum]{revtex4}
\usepackage{graphicx}
\usepackage{dcolumn}
\usepackage[dvipsnames]{xcolor}
\usepackage{bm}
\usepackage{amssymb}
\usepackage[colorlinks=true,linkcolor=WildStrawberry,citecolor 	=WildStrawberry]{hyperref}
\usepackage{amsmath}
\usepackage{tikz}
\usetikzlibrary{decorations.pathmorphing}
\usepackage{subfig}
\usepackage{enumerate}

\newcommand{\be}{\begin{equation}}
\newcommand{\ee}{\end{equation}}
\newcommand{\bea}{\begin{eqnarray}}
\newcommand{\eea}{\end{eqnarray}}
\newcommand{\bmx}{\begin{pmatrix}}
\newcommand{\emx}{\end{pmatrix}}

\usepackage{mathrsfs}

\newcommand{\gra}{\alpha}

\newcommand{\grg}{\gamma}
\newcommand{\grd}{\delta}
\newcommand{\gre}{\epsilon}

\newcommand{\grl}{\lambda}
\newcommand{\grm}{\mu}

\newcommand{\grr}{\rho}
\newcommand{\grs}{\sigma}
\newcommand{\grt}{\tau}

\newcommand{\grf}{\phi}

\newcommand{\grc}{\psi}
\newcommand{\grw}{\omega}

\newcommand{\nn}{\nonumber}

\begin{document}
\title{Asymptotic states of accelerated detectors and universality of the Unruh effect}
\author{Dimitris Moustos}\email{dmoustos@upatras.gr}
\affiliation{Department of Physics, University of Patras, 26504 Patras, Greece}
\date{\today}

\begin{abstract}
We treat an Unruh-DeWitt detector as an open quantum system and evaluate the response of a uniformly accelerated detector: (i) interacting locally with the derivatives of a massless scalar field and (ii) linearly coupled to an electromagnetic field. We find that the early-time transition rate of the detector strongly depends on the type of the interaction between the detector and the quantum field, and may not follow a Planck distribution. In contrast, the late time asymptotic state is always thermal at the Unruh temperature and thus provides a more fundamental and persistent characterization of the acceleration temperature: A uniformly accelerated detector experiences the field vacuum as a {\em genuine thermal bath} at the Unruh temperature and eventually settles at a thermal state, regardless
of their intermediate dynamics or the type of interaction.

\end{abstract}

\maketitle
%=========================================
%=========================================
%=========================================
\section{Introduction}

Quantum field theory (QFT) implies that every inertial observer in Minkowski spacetime agrees on the number of particles in a given field state. However,
this is not the case for noninertial observers. Different noninertial  observers define particles with respect to different field modes \cite{Full, Unruh2, PCDavies}.  The most well-known example is the {\em Unruh effect} \cite{Unruh}: an observer moving with uniform proper acceleration $a$, perceives the  Minkowski vacuum as a heat bath at the Unruh temperature $T_U = a/(2\pi)$. This relation between acceleration and temperature has strong analogies to 
the thermal emission from black holes \cite{hawking} and cosmological horizons \cite{Gibbons}, and as a consequence, Unruh effect constitutes a fundamental ingredient of the theories suggesting a thermodynamic interpretation of gravity \cite{thermgr} (for a comprehensive review see \cite{DM}).

In the framework of QFT, the Unruh effect is usually derived by employing mathematical techniques on nonlocal field modes, as for example Bogoliubov transformations. The derivation also depends on global spacetime properties, such as the existence of Rindler horizons \cite{Full2}. Nevertheless, the Unruh effect can  be expressed in terms of {\em local} physics  employing the notion of the Unruh-DeWitt (UDW) detector \cite{Unruh,Dewitt}. An UDW detector consists of a pointlike quantum system that interacts locally through a monopole coupling with a quantum field and is allowed to move along any trajectory in Minkowski spacetime. The relation between acceleration and temperature is then deduced from the two-point correlation functions of the field in an entirely local way \cite{Unruh, UW84, BL, AM, matsas, review}.

In most discussions of the Unruh effect, the relation between acceleration and temperature is demonstrated by evaluating the excitation rate of a moving UDW detector to leading order in time-dependent perturbation theory \cite{Birrell}. For a uniformly accelerated detector, the excitation rate follows a Planck distribution at the Unruh temperature.
 This feature of the excitation rate is considered as  validation of the Unruh effect. 

However, the perturbative evaluation of the excitation rate has a restricted domain of applicability. It works best for  macroscopic detectors, i.e., systems that leave a macroscopic record every time a particle is detected. For such detectors, the perturbative evaluated transition rate applies at all times, provided that the detector's temporal resolution is sufficiently large \cite{AnSav11}. UDW detectors are not macroscopic particle detectors \cite{paddy}, but rather microscopic {\em field probes}: localized quantum systems---like for example, elementary particles or atoms--- that interact with a quantum field. Information about the field is incorporated in the final state of the probe and is extracted through a suitable measurement \cite{adesso}. In the case of quantum probes, the leading-order perturbative evaluation of the transition rate applies {\em only} during very early times, since it ignores the effect of spontaneous emission after excitations. It also ignores the effect of the the backaction of the field to the detector. In order to take these effects into account, the UDW detectors should be treated as {\em open quantum systems} \cite{Breuer}, with the quantum field playing the role of the environment, inducing dissipation and decoherence.

In a previous work \cite{DMCA}, we evaluated the response of a uniformly accelerated two-level UDW detector linearly coupled to a scalar field in Minkowski spacetime. We treated the detector as an open quantum system and derived the evolution equations of the detector's reduced density matrix, invoking neither the Markov approximation nor the Rotating Wave Approximation (RWA). We demonstrated that the  asymptotic state of the detector is thermal at the Unruh temperature, even when the non-Markovian effects are taken into account. In contrast, the early-time transition rate does not exhibit a thermal behavior when non-Markovian effects are taken into account. The Planckian form of the early-time transition rate is valid only within the Markovian regime, which corresponds to the limit of high accelerations or ultraweak coupling.
 
In this paper, we aim to  examine to what extent the conclusions of our previous work apply for different type of detectors and fields. To this end, we address the response of a uniformly accelerated  detector: (i) coupled to the derivatives of a scalar field and (ii) interacting with an electromagnetic (EM) field.

We find that the early-time transition rate strongly depends on the type of interaction between the detector and the field, and is not Planckian, even in the Markovian regime. In contrast, the asymptotic state of the detector is always thermal at the Unruh temperature, regardless the internal characteristics of the detector or the interacting field. These results reinforce the conclusions made in our previous work \cite{DMCA}: the asymptotic state of an UDW detector provides a more  fundamental and persistent characterization of the acceleration temperature. A uniformly accelerated detector experiences the field vacuum as a  {\em genuine thermal bath} at the Unruh temperature and eventually settles at a state of thermal equilibrium, regardless
of their intermediate dynamics or the type of interaction.

 The structure of the article is the following. In Sec. \ref{rev}, we treat an UDW detector as an open quantum system and derive the time evolution equations of the reduced density matrix of the detector.
In Sec. \ref{UAD}, we solve the evolution equations for a uniformly accelerated UDW detector linearly coupled to a scalar field, reviewing the results of our previous work \cite{DMCA}. In Sec. \ref{dcoupl}, we evaluate the response of a uniformly accelerated detector derivatively coupled to a massless scalar field. In Sec. \ref{EMf}, we address the response of a detector interacting with a quantized EM field. Finally, in Sec. \ref{concl}, we summarize and discuss our results.

 We work with units   $\hbar=c=k_{B}=1$.

%=========================================
%=========================================
%=========================================
\section{Time evolution of UDW detectors}\label{rev}

We model an UDW detector by a two-level system (2LS) of frequency $\grw$. The detector interacts through a monopole coupling with a massless scalar field $\hat{\phi}$ and moves along a trajectory $x^{\mu}(\tau)$ in Minkowski spacetime, where $\tau $ is the proper time of the detector. The Hamiltonian of the combined system is
\be\label{toth}
\hat{H}=\hat{H}_0\otimes\hat{1}+\hat{1}\otimes\hat{H}_{\phi}+\hat{H}_{\text{int}},
\ee
where
\be
\hat{H_0}=\frac{\grw}{2}\hat{\grs}_3
 \ee
 is the 2LS Hamiltonian,
 \be
 \hat{H}_{\grf}=\int d^3x\left(\frac{1}{2}\hat{\pi}^2+\frac{1}{2}(\nabla \hat{\grf})^2\right)
  \ee
  is the Hamiltonian of the scalar field, 
  \be
  \hat{H}_{\text{int}}=g\hat{m}\otimes\hat{\phi}(\mathbf{x})
\ee
is the interaction Hamiltonian,  $g$ is the coupling constant, and $\hat{m} = \hat{\sigma}_1$ is the detector's monopole moment operator. We note that the Hamiltonian
(\ref{toth}) is a special case of the spin-boson Hamiltonian \cite{spinboson}.

The evolution equation of the density matrix $\hat{\grr}_{\text{tot}}$ of the total system in the interaction picture is 
\be\label{von}
\frac{d}{d\grt}\hat{\grr}_{\text{tot}}(\grt)=-i\left[\hat{H}_\text{int}(\grt),\hat{\grr}_{\text{tot}}(\grt)\right],
\ee
where 
\be
\hat{H}_{\text{int}}(\grt)=g\hat{m}(\grt)\otimes\hat{\grf}\left[x^{\grm}(\grt)\right],
\ee
 with the monopole moment
\be\label{mmoment}
\hat{m}(\grt)=e^{i\grw\grt}\hat{\grs}_++e^{-i\grw\grt}\hat{\grs}_-
\ee
being expressed in terms of the SU(2) ladder operators $\hat{\grs}_{\pm}$.

For a weak coupling between the system and the environment,
we solve Eq. (\ref{von}) using the Born approximation: we assume that the state
  of the total system at time $\grt$ approximates a tensor product
\be
\hat{\grr}_{\text{tot}}(\grt)\approx \hat{\grr}(\grt)\otimes\hat{\grr}_{\grf}(0),
\ee
where $\hat{\grr}$ is the reduced density matrix of the 2LS.

Then, Eq. (\ref{von}) becomes an integrodifferential for the reduced density matrix $\hat{\grr}=\text{tr}_{\grf}[\hat{\grr}_{\text{tot}}]$  \cite{Breuer}.
Tracing out the degrees of freedom of the field and assuming that the field is initially in its ground state $\hat{\grr}_{\grf}(0)=|0 \rangle\langle 0|$, we obtain
\bea\label{mastereq}
\dot{\hat{\rho}}(\grt)=g^2\int_{0}^{\grt}d\grt'&&\Big\{\Big(\hat{\grs}_-\hat{\grr}(\grt')\hat{\grs}_-e^{-2i\grw \grt}+\hat{\grs}_-\hat{\grr}(\grt')\hat{\grs}_+
-\hat{\grs}_+\hat{\grs}_-\hat{\grr}(\grt')\Big)e^{i\grw(\grt-\grt')}\nn\\&&+
\Big(\hat{\grs}_+\hat{\grr}(\grt')\hat{\grs}_+e^{2i\grw \grt}+\hat{\grs}_+\hat{\grr}(\grt')\hat{\grs}_-
-\hat{\grs}_-\hat{\grs}_+\hat{\grr}(\grt')\Big)e^{-i\grw(\grt-\grt')}\Big\}\Delta^+(\grt ;\grt')\nn\\&&+ \text{H.c.}
\eea
where $\Delta^+(\grt ;\grt')=\langle 0|\hat{\grf}[x(\grt)]\hat{\grf}[x(\grt')]|0\rangle$ is the positive-frequency Wightman function and $\Delta^-(\grt ;\grt')=\langle 0|\hat{\grf}[x(\grt')]\hat{\grf}[x(\grt)]|0\rangle$ is the negative-frequency Wightman  function. Equation (\ref{mastereq}) is a non-Markovian time evolution equation: the evolution of the $\hat{\rho}(\grt)$ depends on its past history (memory effects) through the integration over $\hat{\grr}(\grt')$. 
Thus, Eq. (\ref{mastereq}) is valid at all times,  for any trajectory followed by the detector and takes into account the backaction of the field to the detector. It is derived using only the Born approximation. We  used neither the  Markov approximation (which neglects the memory effects) nor the (post-trace) RWA \cite{RWA} (where rapidly oscillating terms  in the interaction picture evolution equation are ignored).

For stationary Wightman functions, i.e.,  $\Delta^{\pm} (\grt; \grt') = \Delta^{\pm} (\grt- \grt')$, Eq. (\ref{von}) is easily solved employing the Laplace transform method and the convolution theorem \cite{DMCA, Davies}. 
This is possible for a specific class of spacetime trajectories \cite{LePf, Letaw}, which includes, for example, trajectories with constant proper acceleration and with rotation at constant angular velocity.

\section{Uniformly accelerated detectors}\label{UAD}

For a uniformly accelerated detector following the hyperbolic trajectory
\be\label{atraj}
x^{\grm}(\grt)=\left(a^{-1}\sinh(a\grt),a^{-1}\cosh(a\grt),0,0\right),
\ee
where $a$ is the proper acceleration, the  corresponding Wightman functions are
\be\label{Wacc}
\Delta^{\pm}(\grt -\grt')= -\lim_{\epsilon \rightarrow 0^+} \frac{a^2}{16\pi^2\sinh^2[a(\grt-\grt' \mp i\gre)/2]}.
\ee

For the stationary Wightman function (\ref{Wacc}), Eq. (\ref{von}) are analytically solved  (for more details see \cite{DMCA}), and we obtain
for the reduced density matrix elements 
\begin{eqnarray}
\grr_{11}(\grt)&=&\frac{1}{2}\left(1-\frac{\Gamma_0}{\Gamma}\right)+\frac{\Gamma_0}{2\Gamma}e^{-\Gamma \grt}-\frac{1}{2}\left[e^{-\Gamma \grt}+\frac{2\Gamma_0}{ \grw}S_1(\grw;a;\grt)\right](\grr_{00}(0)-\grr_{11}(0)), \label{r11}\\
\grr_{00}(\grt) &=& 1 - \grr_{11}(\grt), \label{r00}\\
\grr_{10}(\grt)&=& \frac{\omega}{\bar{\omega}} e^{-i \bar{\omega}\grt -\frac{\Gamma}{2}\grt}\grr_{10}(0) - \frac{\Gamma_0}{\omega} S_2(\grw;a;\grt)\grr_{10}(0) + \frac{2\Gamma_0}{\omega} S_3(\grw;a;\grt)\grr_{10}(0) - \frac{\Gamma_0}{\omega} S_2(\grw;a;\grt)\grr_{01}(0) 
\nn \\
&& - \frac{\bar{\omega}-\omega}{\bar{\omega}}  e^{-\frac{\Gamma}{2} \grt}\cos(\bar{\omega} \tau) \grr_{01}(0) +\frac{\Gamma}{2 \bar{\omega}}e^{-\frac{\Gamma}{2} \grt} \sin(\bar{\omega} \tau)\grr_{01}(0) \label{r10}.
\end{eqnarray}
In the above expressions,
\begin{eqnarray}
 \Gamma =  \Gamma_0   \coth\left(\frac{\pi\grw}{a}\right) \label{decaya}
\end{eqnarray}
is a thermal decay constant, expressed in terms of the decay constant of a static 2LS
\be
\Gamma_0 = \frac{g^2\grw}{2\pi}.
\ee
We have defined a shifted frequency as 
\be\label{ombar}
\bar{\omega} = \omega + \left(C_R + \Delta \omega\right),
\ee
where
\be\label{renorm}
C_R = \frac{\Gamma_0}{\pi}\log(e^{\grg}\gre\grw)
\ee
is an acceleration independent frequency renormalization term that diverges logarithmically as $\epsilon \rightarrow 0^+$, and
\be\label{Lamb}
\Delta\grw=\frac{\Gamma_0}{\pi}\left[\log(a/\grw )+Re\left\{\grc\left(\frac{i\grw}{a}\right)\right\}\right]
\ee
is the finite Lamb-shift of the frequency due to acceleration; $\psi(z)$ is the digamma (psi) function. The frequency shift  (\ref{Lamb}) vanishes when $a\to0$, since $\lim_{z\to0}\psi(z)\sim \log(z)$. 

The divergent renormalization term $C$ added to the bare frequency $\grw$ appears due to the interaction of the system with the infinite modes of the environment. We absorb this divergent term into a redefinition of the two-level system's frequency. 
In a complete QFT treatment of the total system, one needs to include a second-order to the coupling counter term in the total Hamiltonian (\ref{toth}), in order to compensate for the renormalization \cite{spinboson,Breuer}. 

The functions $S_1(\grw;a;\grt), S_2(\grw;a;\grt)$ and $S_3(\grw;a;\grt)$ incorporates the non-Markovian effects. They do not appear in the solution of Eq. (\ref{mastereq}), when the Markov approximation is employed. To leading-order in $g^2$ they read
\bea
S_1(\grw;a;\grt)&=&\frac{1}{\pi}Re\left\{\sum_{n=1}^{\infty}\frac{ne^{-na\grt}}{\left(n-i\frac{\grw}{a}\right)^2}\right\},\label{S1}
\eea
\be\label{S2}
S_2(\grw;a;\grt)=\frac{1}{\pi} \sum_{n=1}^{\infty}e^{-na\grt}\frac{n}{n^2+(\frac{\grw}{a})^2}\quad \text{and} \quad S_3(\grw;a;\grt)= \frac{1}{\pi} \sum_{n=1}^{\infty}e^{-na\grt}\frac{n^2(n+i \grw/a)}{\left(n^2+(\frac{\grw}{a})^2\right)^2}.
\ee

\begin{figure}[t!]
\subfloat[$\grw/a=1.1$]{\includegraphics[scale=0.46]{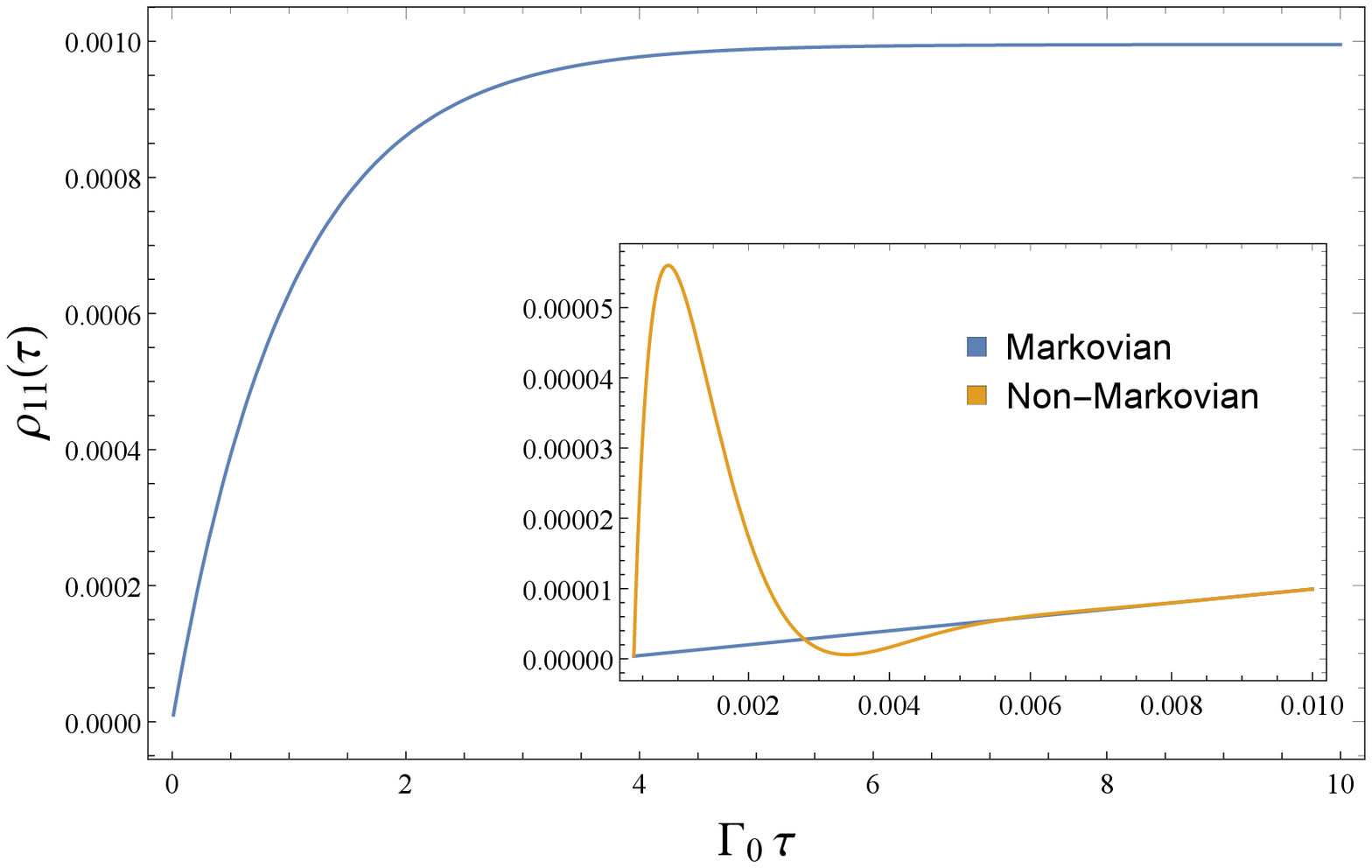}}  \hspace{0.2cm}
\subfloat[$\grw/a=0.8$]{\includegraphics[scale=0.46]{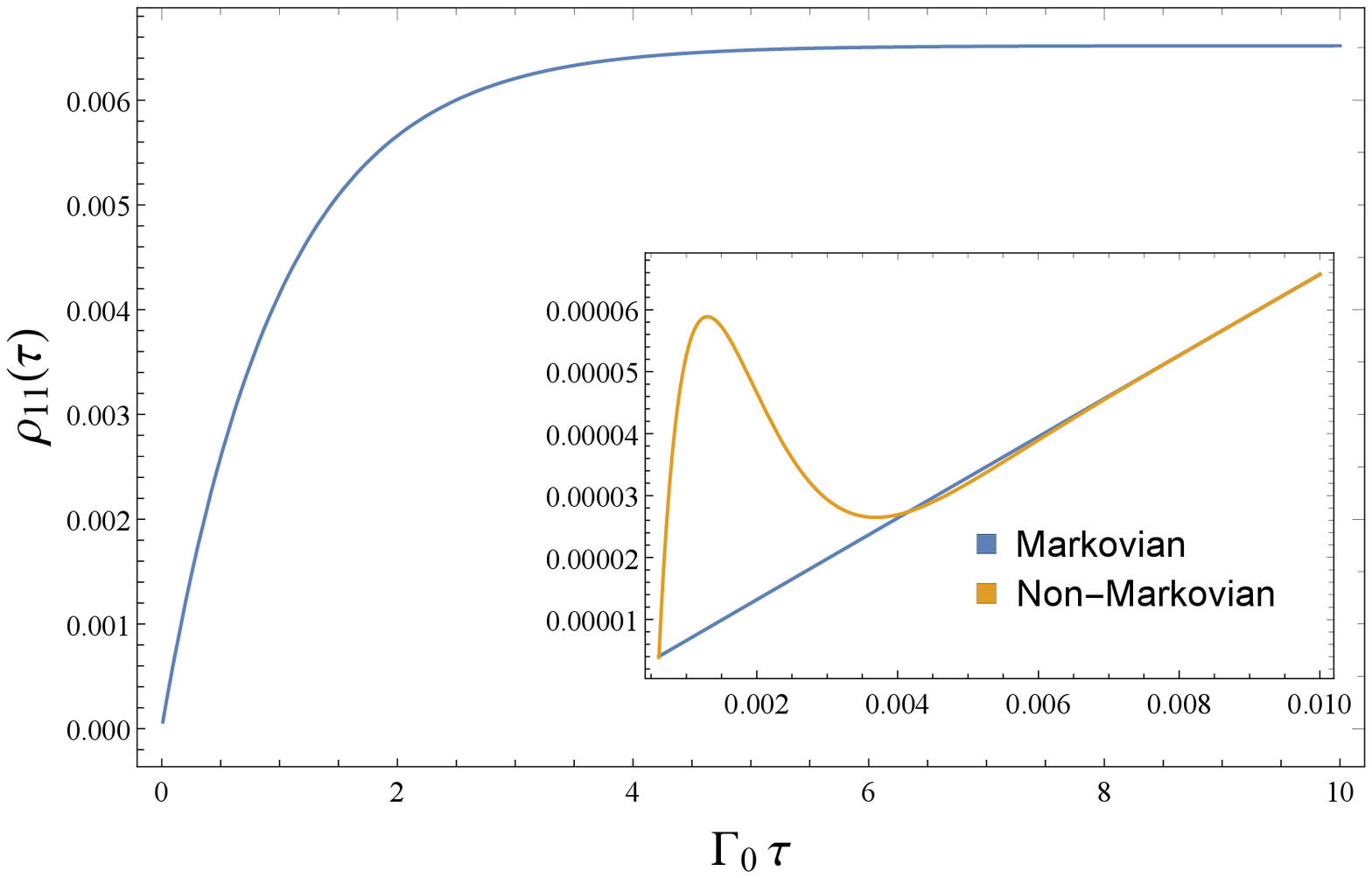}} \hspace{1cm}
\subfloat[$\grw/a=0.6$]{\includegraphics[scale=0.46]{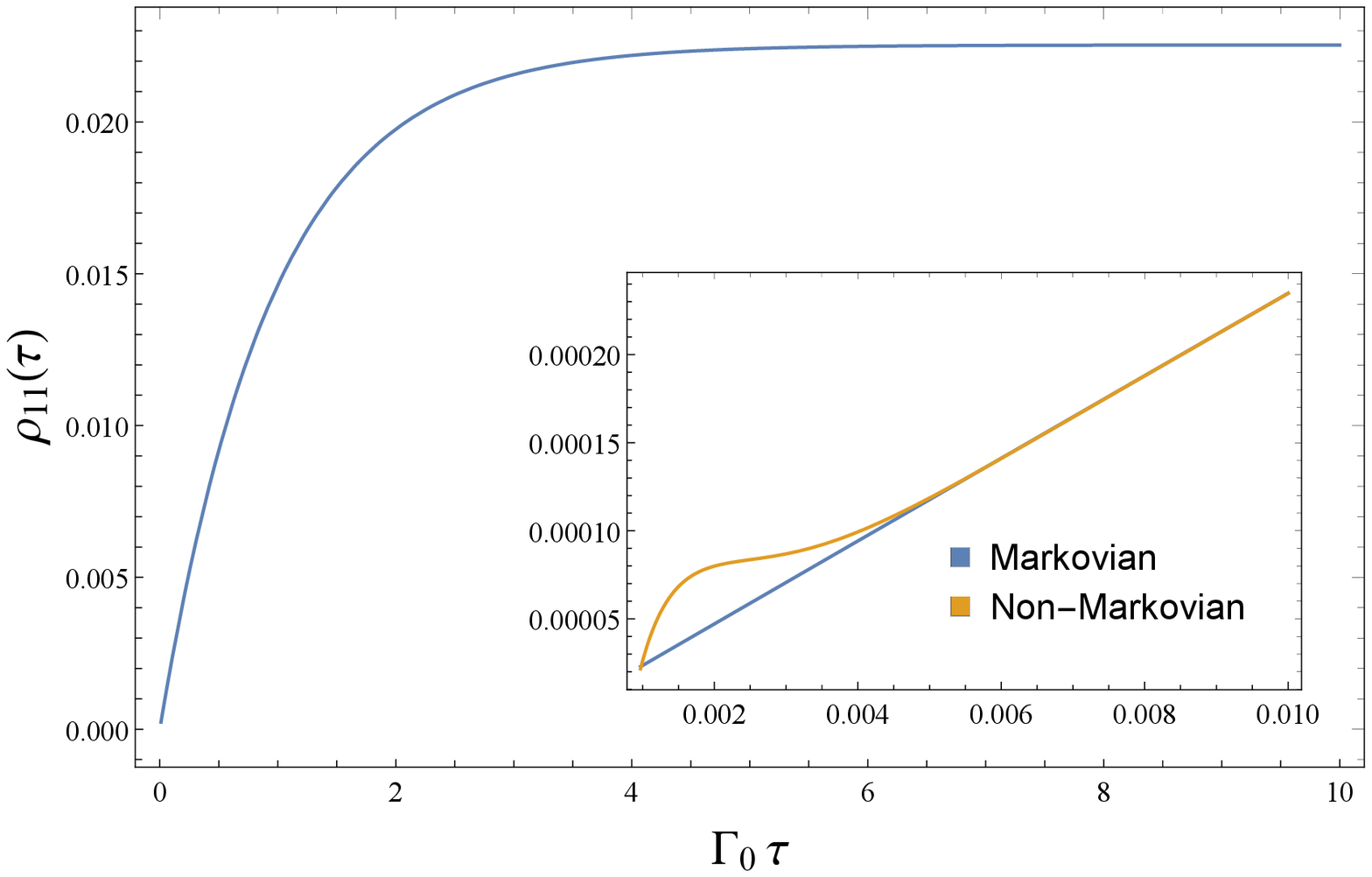}} \hspace{0.2cm}
\subfloat[$\grw/a=0.4$]{\includegraphics[scale=0.46]{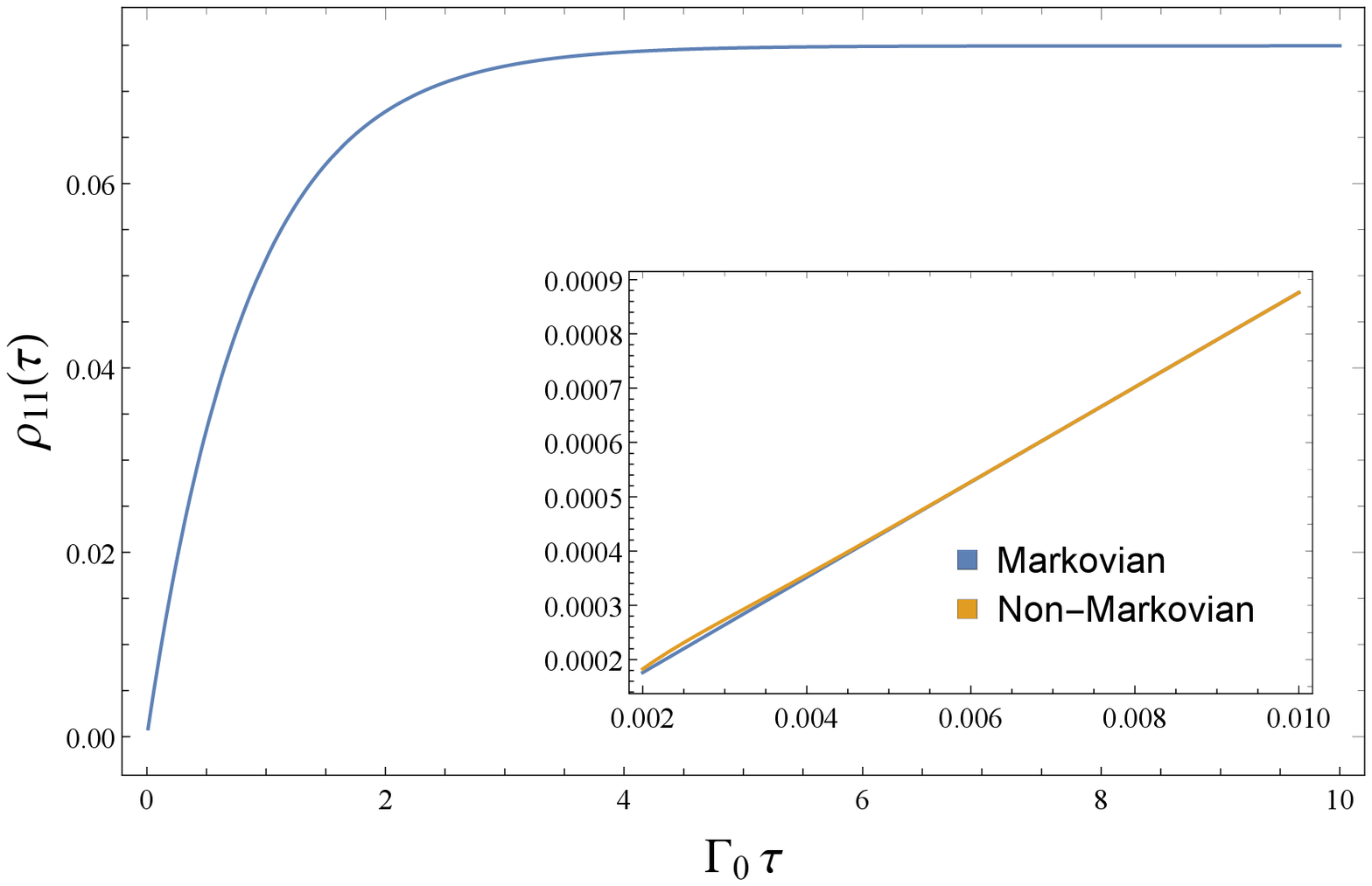}} 
\caption{Time evolution of the $\grr_{11}(\grt)$ element (\ref{r11}) of the detector's density matrix for different values of $\grw/a$. The detector is initially found in its ground state $\grr_{00}(0)=1$. Evolution at early times is shown in the inserted plots.}
\label{scalar}
\end{figure}

\subsection{Implications}

At early times ($\grt$ of order $\omega^{-1}$)  and for $a$ close to $\omega$ or smaller, the non-Markovian terms induce strong oscillations and dominate over the Markovian terms \cite{DMCA}. This behavior is depicted in Fig. \ref{scalar}. At later times and for $\alpha < \Gamma$ convergence to equilibrium is also dominated by the non-Markovian terms, and the relaxation time  is of the order of  $a^{-1}$ rather than $\Gamma^{-1}$. Non-Markovian effects become absent in the regime of high accelerations $a>\grw$ \cite{DMCA}. 

In the Markovian regime of high accelerations, the term $S_1(\grw;a;\grt)$ in (\ref{r11}) vanishes. Then, at early times ($\Gamma t << 1$), and for $\rho_{11}(0) = 0$, Eq. (\ref{r11}) implies that
\be
\grr_{11}(\grt) =\Gamma_0\frac{\grt}{e^{\frac{2\pi\grw}{a}}-1} \label{thrmtr}
\ee
and the transition rate (transition probability per unit time) to the excited state is 
\begin{eqnarray}\label{rate}
w = \frac{\Gamma_0}{e^{\frac{2\pi\grw}{a}}-1}. \label{trr}
 \label{tranw}
 \end{eqnarray}
This is identical to the transition  rate of  a static 2LS in a thermal bath at the Unruh temperature. When $a\to 0$, the transition
rate vanishes. There are no excitations in the case of an inertial UDW detector. The rate (\ref{rate}) coincides with the rate commonly obtained through first order perturbation theory \cite{Dewitt, Birrell} and is invoked in most discussions of acceleration temperature.
For $a< \omega$, non-Markovian effects imply a non-Planckian transition rate.

In the long-time limit ($\Gamma t >> 1$), the density matrix approaches the equilibrium value
\be
\hat{\rho}_{\infty} = \left( \begin{array}{cc} \frac{e^{-\frac{2\pi\grw}{a}}}{e^{-\frac{2\pi\grw}{a}}+1}& 0 \\0 &\frac{1}{e^{-\frac{2\pi\grw}{a}}+1} \end{array} \right), \label{asympt}
\ee
which is a Gibbs thermal state at the Unruh temperature $T_U = a/(2\pi)$. Equation (\ref{asympt}) applies even when we keep the contribution of the non-Markovian terms. The accelerated detector experiences the field vacuum as a genuine thermal bath and eventually settles at a thermal state.

%=========================================
%=========================================
%=========================================
\section{Derivative coupling detector}\label{dcoupl}
We consider a 2LS detector of frequency $\grw$ interacting with the derivative of a massless scalar field $\hat{\grf}$ \cite{Hinton, Grove, ZhuYu07}. The interaction Hamiltonian in the interaction picture reads
\be
\hat{H}_{\text{int}}(\grt)=g\hat{m}^{\grm}(\grt)\otimes\partial_{\grm}\hat{\grf}\left[x(\grt)\right],
\ee
where $g$ is a coupling constant and $\partial_{\grm}=\partial/\partial x^{\grm}$. The dipole moment operator $\hat{m}^{\grm}(\grt)$ can be orientated in different ways. 

Next, we study the following alternatives of the derivative coupling UDW detector: (i) a detector coupled to the proper time derivative of the scalar field, (ii) a detector coupled to the derivative of the spatial component parallel to the direction of acceleration and (iii) a detector coupled to the derivative of the spatial component perpendicular to the direction of acceleration. We examine the above situations both in the two-dimensional and four-dimensional Minkowski spacetime.

%=========================================
%=========================================
%=========================================
\subsection{(1+1)-dimensional Minkowski spacetime}

For a massive quantum scalar field with mass $m$ in (1+1) dimensions the Wigthman function reads \cite{Birrell, Takagi}
\be
\Delta^{\pm}(\grt-\grt')=\frac{1}{2\pi}K_0\left(m\sqrt{-(\grt-\grt'\mp i\gre)^2+({\bf x-x'})^2}\right),
\ee
where $K_0$ is the modified Bessel function of the second kind \cite{Ryzhik} and $\gre\to 0^+$. Taking the massless limit $m\to0$ we obtain
\be
\Delta^{\pm}(\grt-\grt')=-\frac{1}{2\pi}\log{\left(\sqrt{-(\grt-\grt'\mp i\gre)^2+({\bf x-x'})^2}\right)}-\frac{1}{2\pi}\lim_{m\to 0}\log\left(\frac{e^{\grg}m}{2}\right)\label{massW},
\ee
where $\grg$ is the Euler-Mascheroni constant. We note that the Wightman function (\ref{massW}) for a massive scalar field diverges as $m\to0$.  To deal with these infrared ambiguities of the correlation functions, a  detector coupled to the proper time derivative of a scalar field was employed in \cite{Benito}.

When the detector is coupled to the derivative of a scalar field the Wightman functions for a uniformly accelerated detector become
\be\label{11Wdcoupl}
\partial_{\grt}\partial_{\grt'}\Delta^{\pm}(\grt-\grt')=\partial_{x}\partial_{x'}\Delta^{\pm}(\grt-\grt')=-\lim_{\gre\to 0^+}\frac{a^2}{8\pi}\frac{1}{\sinh^2[a(\grt-\grt'\mp i\gre)/2]},
\ee 
where $x$ stands for the spatial component. The correlation functions (\ref{11Wdcoupl}) for a uniformly accelerated derivative coupling detector has the same form with the Wigthman functions (\ref{Wacc}).

Thus, the basic results of Sec. \ref{UAD} are straightforwardly applied in the case of a uniformly accelerated detector derivatively coupled to a massless scalar field in (1+1)-dimensional Minkowski spacetime: 
the transition rate to the excited state within the Markovian regime is
\be
w=2\pi\Gamma_{0}\frac{1}{e^{\frac{2\pi\grw}{a}}-1},
\ee
where $2\pi\Gamma_0=g^2\grw$ is the decay constant of a static derivative coupling detector. The transition rate follows a Planck distribution at the Unruh temperature $T_U$. However, when the full non-Markovian solution of the detector's reduced density matrix is considered, the transition rate at early times is non-Planckian.
On the contrary, the late time asymptotic state of the detector is thermal at the Unruh temperature even when non-Markovian effects are taken into account.

%=========================================
%=========================================
%=========================================
\subsection{(3+1)-dimensional Minkowski spacetime}

In Sec. \ref{UAD}, we demonstrated that in the regime of small accelerations (or equivalently high frequencies) $a<\grw$, non-Markovian effects are particularly pronounced and render the early-time transition rate non-Planckian. Similar to $S_1(\grw;a;\grt), S_2(\grw;a;\grt)$ and $S_3(\grw;a;\grt)$ non-Markovian terms in Eq. (\ref{r11})--(\ref{r10}) appear also in the case of the derivative coupling detector, affecting its evolution in exactly the same way (as illustrated in Fig. \ref{scalar}). The only difference is the form of the expressions for the emission rates $\Gamma_0$ and $\Gamma$. To find these expressions we focus, from now on, on the Markovian time evolution of the derivate coupling  detector.

The Markov approximation is obtained, if in Eq. (\ref{mastereq}) we replace the density matrix $\hat{\rho}(s)$ by $\hat{\rho}(\grt)$ and take the upper limit to infinity \cite{Breuer, vega}. We note that the Markov approximation is justified only if the correlation functions of the environment decay very rapidly compared to the time scale on which $\hat{\rho}(\grt)$ changes. Expressing the density operator in a matrix form
\be\label{dmatrix}
\hat{\grr}(\grt)=
\bmx
\grr_{11}(\grt)&\grr_{10}(\grt)\\
\grr_{01}(\grt)&\grr_{00}(\grt)
\emx
\ee
and performing the Markov approximation we obtain
\bea\label{mr11}
\dot{\grr}_{11}(\grt) =&-&g^2\grr_{11}(\grt)\int_0^{\infty}dt\left[e^{i\grw t}W^{+}(t)+e^{-i\grw t}W^{-}(t)\right]
\nonumber \\
&+& g^2\grr_{00}(\grt)\int_0^{\infty}dt\left[e^{-i\grw t}W^{+}(t)+e^{i\grw t}W^{-}(t)\right],
\eea
\bea\label{mr00}
\dot{\grr}_{00}(\grt) =&-&g^2\grr_{00}(\grt)\int_0^{\infty}dte^{-i\grw t}\left[W^{+}(t)+e^{i\grw t}W^{-}(t)\right] \nonumber \\
&+&g^2\grr_{11}(\grt)\int_0^{\infty}dt\left[e^{i\grw t}W^{+}(t)+e^{-i\grw t}W^{-}(t)\right],  
\eea
\bea \label{mr10}
\dot{\grr}_{10}(\grt)=&-&g^2\grr_{10}(\grt)\int_0^{\infty}dt\left[e^{i\grw t}W^{+}(t)+e^{i\grw t}W^{-}(t)\right]\nn\\&+&g^2\grr_{01}(\grt)e^{2i\grw\grt}\int_0^{\infty}dt\left[e^{-i\grw t}W^{-}(t)+e^{-i\grw t}W^{+}(t)\right].
\eea
In Eqs. (\ref{mr11})--(\ref{mr10}) we represent by $W$ the Wightman functions of a derivative coupling detector.

For the various cases of the derivative coupling detector presented in the following sections, the Laplace transformlike integrals involved in Eqs.  (\ref{mr11})--(\ref{mr10}), are analytically evaluated in the Appendix \ref{LPT}.

\subsubsection{Proper time derivative}

We first consider a  detector coupled to the proper time derivative of a massless scalar field. The detector follows the trajectory (\ref{atraj}). The corresponding Wightman functions reads
\be\label{Wtd}
W_{\text{pt}}^{\pm}(\grt-\grt')=\partial_{\grt}\partial_{\grt'}\Delta^{\pm}(\grt-\grt')=\frac{3a^4}{32\pi^2}
\frac{1}{\sinh^4\left[a(\grt-\grt'-\mp i\gre)/2\right]}+\frac{a^4}{16\pi^2}
\frac{1}{\sinh^2\left[a(\grt-\grt'-\mp i\gre)/2\right]}
\ee
Inserting the correlation functions (\ref{Wtd}) into the evolution equations (\ref{mr11}--\ref{mr10}), we evaluate the integral transforms of the Wightman functions (see the Appendix \ref{LPT}) and solve the set of equations to obtain
\begin{eqnarray}
\grr_{11}(\grt)&=&\frac{1}{2}\left(1-\frac{\Gamma^{\text{pt}}_0}{\Gamma^{\text{pt}}}\right)+\frac{\Gamma^{\text{pt}}_0}{2\Gamma^{\text{pt}}}e^{-\Gamma^{\text{pt}} \grt}-\frac{1}{2}e^{-\Gamma^{\text{pt}} \grt}(\grr_{00}(0)-\grr_{11}(0)),\label{pt11} \\
\grr_{00}(\grt) &=& 1 - \grr_{11}(\grt),\\
\grr_{10}(\grt)&=& \frac{\omega}{\bar{\omega}} e^{-i \bar{\omega}\grt -\frac{\Gamma^{\text{pt}}}{2}\grt}\grr_{10}(0)  
- \frac{\bar{\omega}-\omega}{\bar{\omega}}  e^{-\frac{\Gamma^{\text{pt}}}{2} \grt}\cos(\bar{\omega} \tau) \grr_{01}(0) +\frac{\Gamma^{\text{pt}}}{2 \bar{\omega}}e^{-\frac{\Gamma^{\text{pt}}}{2} \grt} \sin(\bar{\omega} \tau)\grr_{01}(0). \label{offdiag}
\end{eqnarray}
In the above expressions,
\begin{eqnarray}
 \Gamma^{\text{pt}} &=&-\frac{g^2\grw^2 a}{2\pi^2} \bigg\{1+\frac{i\grw}{a}\left[\grc\left(\frac{i\grw}{a}\right)-\grc\left(-\frac{i\grw}{a}\right)\right]\bigg\}\nn\\
&=& \Gamma^{\text{pt}}_0\coth\left(\frac{\pi\grw}{a}\right)
 \label{decaya}
\end{eqnarray}
is a thermal decay constant and 
\be
\Gamma^{\text{pt}}_0=\frac{g^2\grw^3 }{2\pi}
\ee
is the decay constant in the static case. To calculate  $\Gamma^{\text{pt}}$ we used the relation $\grc(z)-\grc(-z)=-1/z-\pi\cot(\pi z)$ for the digamma functions $\grc(z)$. We also defined the shifted frequency 
\be\label{otn}
\bar{\grw}=\grw+\grw^2\left(C_R+\Delta \grw\right)+\frac{\Gamma^{\text{pt}}_0}{\pi}C'_R,
\ee
where $C_R$ is the renormalization term (\ref{renorm}) and $\Delta \grw$ the Lamb shift due to acceleration (\ref{Lamb}).
Compared to (\ref{ombar}), in (\ref{otn}) appears an extra divergent term $C'_R=(a/2)\sin^{-2}(\gre a/2)$ that depends on the acceleration. To compensate for this infinite term, a suitable counter-term should be added to the total Hamiltonian. 

At early times ($\Gamma t << 1$), and for $\rho_{11}(0) = 0$, the transition rate to the excited state reads
\be
w_{pt}=\frac{\Gamma^{\text{pt}}_0}{e^{\frac{2\pi\grw}{a}}-1}.
\ee
The transition rate follows a Planck spectrum at the Unruh temperature $T_U$.

\subsubsection{Direction parallel to acceleration}
We next consider a  detector coupled to the derivative of the spatial component parallel to the direction of acceleration. The correlation functions are \cite{Takagi}
\bea\label{wxx}
W_{\parallel}^{\pm}(\grt-\grt')=\partial_{x}\partial_{x'}\Delta^{\pm}(\grt-\grt')&=&\frac{a^4}{32\pi^2}
\frac{1}{\sinh^4\left[a(\grt-\grt'-\mp i\gre)/2\right]}-\frac{a^4}{16\pi^2}
\frac{1}{\sinh^2\left[a(\grt-\grt'-\mp i\gre)/2\right]}\nn\\
&=&\frac{1}{3}\partial_{\grt}\partial_{\grt'}\Delta^{\pm}(\grt-\grt')-
\frac{4a^2}{3}\Delta^{\pm}(\grt-\grt'),
\eea
where $x$ is the spatial direction parallel to acceleration and $\Delta^{\pm}(\grt-\grt')$ is given in (\ref{Wacc}). In the second line of (\ref{wxx}), we write the correlation functions in an appropriate form so that we can exploit the calculations of the Laplace transform-like integrals in the Appendix \ref{LPT}.

Inserting the correlation functions (\ref{wxx}) into Eqs. (\ref{mr11})--(\ref{mr10}), we solve the set of equations to obtain
\begin{eqnarray}
\grr_{11}(\grt)&=&\frac{1}{2}\left[1-\frac{\Gamma^{\parallel}_0 }{\Gamma^{\parallel} }\left(1+\frac{4a^2}{\grw^2}\right)\right]+\frac{\Gamma^{\parallel}_0}{2\Gamma^{\parallel} }\left(1+\frac{4a^2}{\grw^2}\right)e^{-\Gamma^{\parallel}  \grt}-\frac{1}{2}e^{-\Gamma^{\parallel}  \grt}(\grr_{00}(0)-\grr_{11}(0)),\label{par11} \\
\grr_{00}(\grt) &=& 1 - \grr_{11}(\grt), \\
\grr_{10}(\grt)&=& \frac{\omega}{\bar{\omega}} e^{-i \bar{\omega}\grt -\frac{\Gamma^{\parallel}}{2}\grt}\grr_{10}(0)  
- \frac{\bar{\omega}-\omega}{\bar{\omega}}  e^{-\frac{\Gamma^{\parallel}}{2} \grt}\cos(\bar{\omega} \tau) \grr_{01}(0) +\frac{\Gamma^{\parallel}}{2 \bar{\omega}}e^{-\frac{\Gamma^{\parallel}}{2} \grt} \sin(\bar{\omega} \tau)\grr_{01}(0),
\end{eqnarray}
where the thermal decay constant
 \begin{eqnarray}
 \Gamma^{\parallel} = \Gamma^{\parallel}_0 \left(1+\frac{4a^2}{\grw^2}\right)  \coth\left(\frac{\pi\grw}{a}\right) \label{decaya}
\end{eqnarray}
is expressed in terms of the decay constant  in the static case
\be
\Gamma^{\parallel}_0 = \frac{g^2\grw^3}{6\pi}.
\ee
The shifted frequency  is
\be
\bar{\grw}=\grw+\frac{\grw^2}{3}\left(1+\frac{4a^2}{\grw^2}\right)\left(C_R+\Delta \grw\right)+\frac{\Gamma^{\parallel}_0}{\pi}C'_R.
\ee

Evaluating the early-time transition rate for $\rho_{11}(0) = 0$, we obtain
\be\label{xxrate}
w_{\parallel}=\Gamma^{\parallel}_0\left(1+\frac{4a^2}{\grw^2}\right)\frac{1}{e^{\frac{2\pi\grw}{a}}-1}.
\ee
The transition rate (\ref{xxrate}) is non-Planckian. A Planck spectrum is obtained either when the acceleration $a$ is about equal to the detector's frequency $\grw$ or in the regime of small accelerations $a \ll \grw$. In this regimes, however, we have demonstrated that the non-Markovian effects are really significant, rendering the transition rate non-thermal \cite{DMCA}.

\subsubsection{Direction perpendicular to acceleration}

We finally consider a  detector coupled to the derivative of the spatial component perpendicular to the direction of acceleration. The Wightman functions are \cite{Takagi}
\bea\label{wperp}
W_{\bot}^{\pm}(\grt-\grt')&=&\partial_{y}\partial_{y'}\Delta^{\pm}(\grt-\grt')=\partial_{z}\partial_{z'}\Delta^{\pm}(\grt-\grt')\nn\\&=&\frac{a^4}{32\pi^2}
\frac{1}{\sinh^4\left[a(\grt-\grt'-\mp i\gre)/2\right]}\nn\\
&=&\frac{1}{3}\partial_{\grt}\partial_{\grt'}\Delta^{\pm}(\grt-\grt')+
\frac{a^2}{3}\Delta^{\pm}(\grt-\grt')
\eea

Inserting the correlation functions (\ref{wperp}) into Eqs. (\ref{mr11})--(\ref{mr10}) , we solve the set of equations to obtain
\begin{eqnarray}
\grr_{11}(\grt)&=&\frac{1}{2}\left[1-\frac{\Gamma^{\bot}_0}{\Gamma^{\bot}}\left(1+\frac{a^2}{\grw^2}\right) \right]+\frac{\Gamma^{\bot}_0}{2\Gamma^{\bot}}\left(1+\frac{a^2}{\grw^2}\right) e^{-\Gamma^{\bot} \grt}-\frac{1}{2}e^{-\Gamma^{\bot} \grt}(\grr_{00}(0)-\grr_{11}(0)), \label{perp11} \\
\grr_{00}(\grt) &=& 1 - \grr_{11}(\grt)\nn\\
\grr_{10}(\grt)&=& \frac{\omega}{\bar{\omega}} e^{-i \bar{\omega}\grt -\frac{\Gamma^{\bot}}{2}\grt}\grr_{10}(0),  
- \frac{\bar{\omega}-\omega}{\bar{\omega}}  e^{-\frac{\Gamma^{\bot}}{2} \grt}\cos(\bar{\omega} \tau) \grr_{01}(0) +\frac{\Gamma^{\bot}}{2 \bar{\omega}}e^{-\frac{\Gamma^{\bot}}{2} \grt} \sin(\bar{\omega} \tau)\grr_{01}(0),
\end{eqnarray}
where the thermal decay constant reads
\begin{eqnarray}
 \Gamma^{\bot} =  \Gamma^{\bot}_0 \left(1+\frac{a^2}{\grw^2}\right)  \coth\left(\frac{\pi\grw}{a}\right) \label{decaya}
\end{eqnarray}
and the decay constant in the static case is 
\be
\Gamma^{\bot}_0 = \frac{g^2\grw^3}{6\pi}.
\ee
The shifted frequency  is
\be
\bar{\grw}=\grw+\frac{\grw^2}{3}\left(1+\frac{a^2}{\grw^2}\right)\left(C_R+\Delta \grw\right)+\frac{\Gamma^{\bot}_0}{\pi}C'_R
\ee

Evaluating the early-time transition rate for $\rho_{11}(0) = 0$, we obtain
\be\
w_{\bot}=\Gamma^{\bot}_0\left(1+\frac{a^2}{\grw^2}\right)\frac{1}{e^{\frac{2\pi\grw}{a}}-1}.
\ee
Again, the transition rate is non-Planckian, except the regimes where the acceleration is equal or much smaller than the detector's frequency. In these regimes, however, non-Markovian effects are significant and cannot be neglected.

\subsubsection{Asymptotic states and thermal behavior}

We demonstrated that in the cases of a uniformly accelerated detector coupled to the spatial derivatives of a scalar field, the early-time transition rate to an excited state does not obey a Planck distribution, even in the Markovian regime of high accelerations. However, the thermality of the Unruh effect should not be identified with the Planckian form of the transition rate. In every alternative of the derivative coupling detector presented previously and in the long-time limit ($\Gamma t >> 1$), the density matrix of the detector approaches the equilibrium value
\be
\hat{\rho}_{\infty} = \left( \begin{array}{cc} \frac{e^{-\frac{2\pi\grw}{a}}}{e^{-\frac{2\pi\grw}{a}}+1}& 0 \\0 &\frac{1}{e^{-\frac{2\pi\grw}{a}}+1} \end{array} \right), 
\ee
which is a thermal state at the Unruh temperature $T_U=a/(2\pi)$. The asymptotic state is thermal even if we keep the contribution of the non-Markovian terms in the solutions of the evolution equations.
This is a much stronger manifestation of the thermal behavior than the transition rate. The accelerated
detector experiences the field vacuum as a {\em genuine  thermal bath} and eventually settles at a thermal state. 

\begin{figure}[t!]
\subfloat[$\grw/a=0.2$]{\includegraphics[scale=0.61]{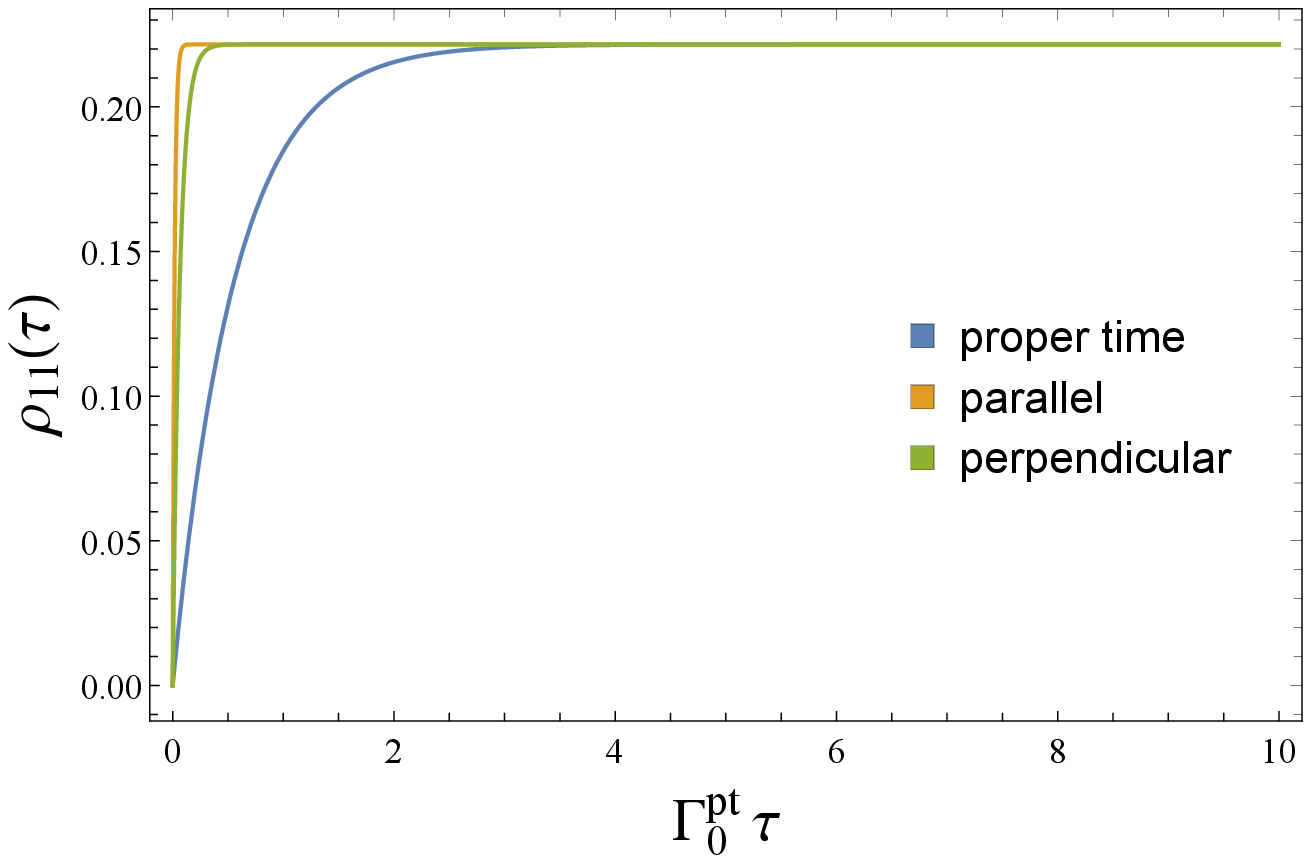}}  \hspace{0.2cm}
\subfloat[$\grw/a=0.5$]{\includegraphics[scale=0.61]{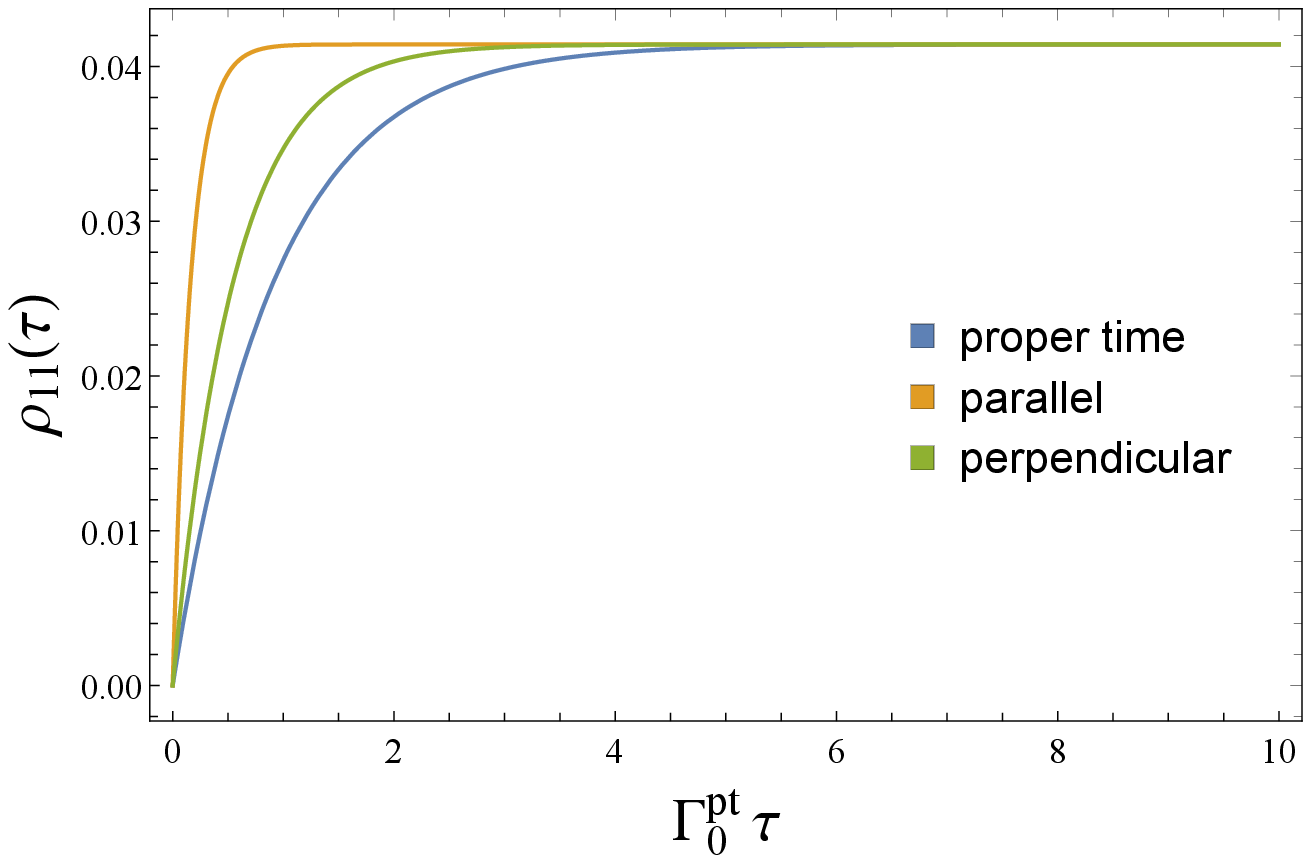}} \hspace{1cm}
\subfloat[$\grw/a=1$]{\includegraphics[scale=0.61]{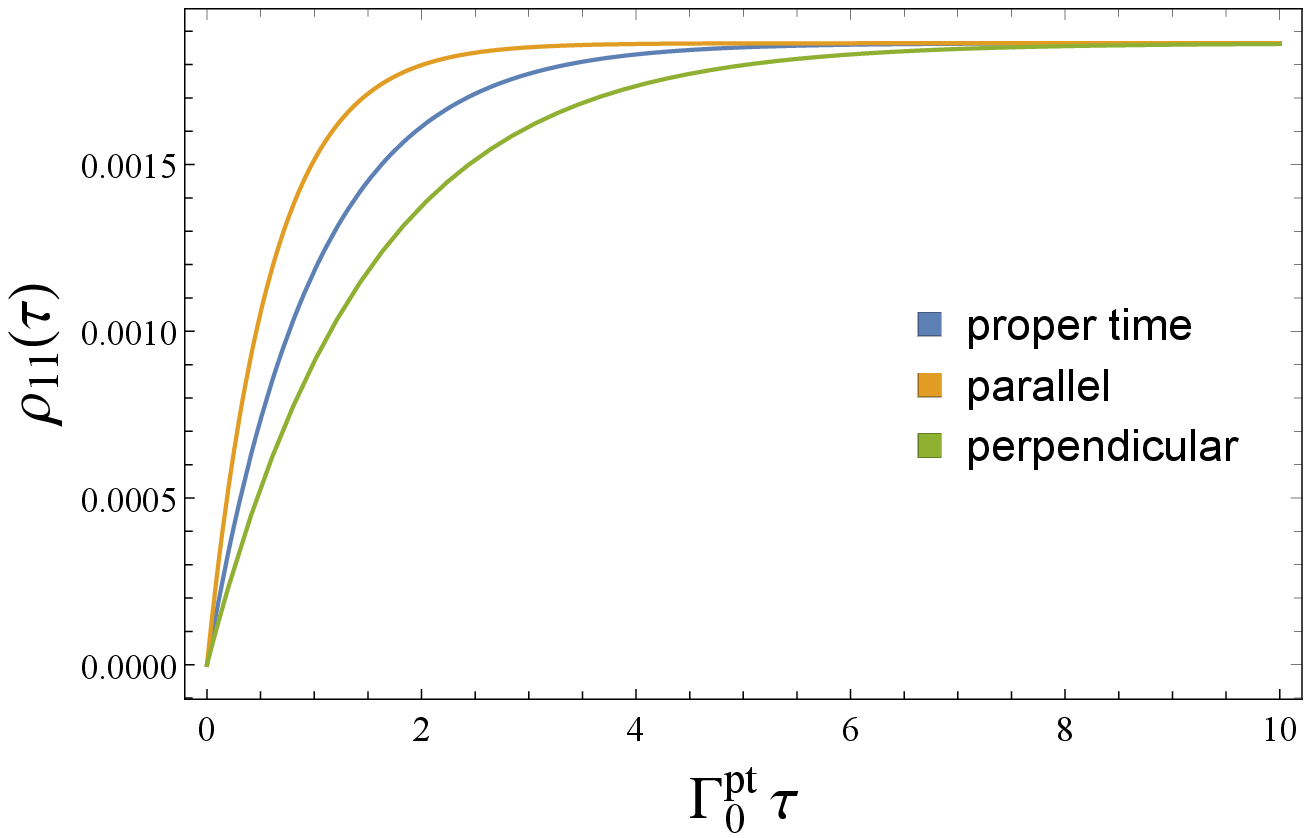}} \hspace{0.2cm}
\subfloat[$\grw/a=1.6$]{\includegraphics[scale=0.61]{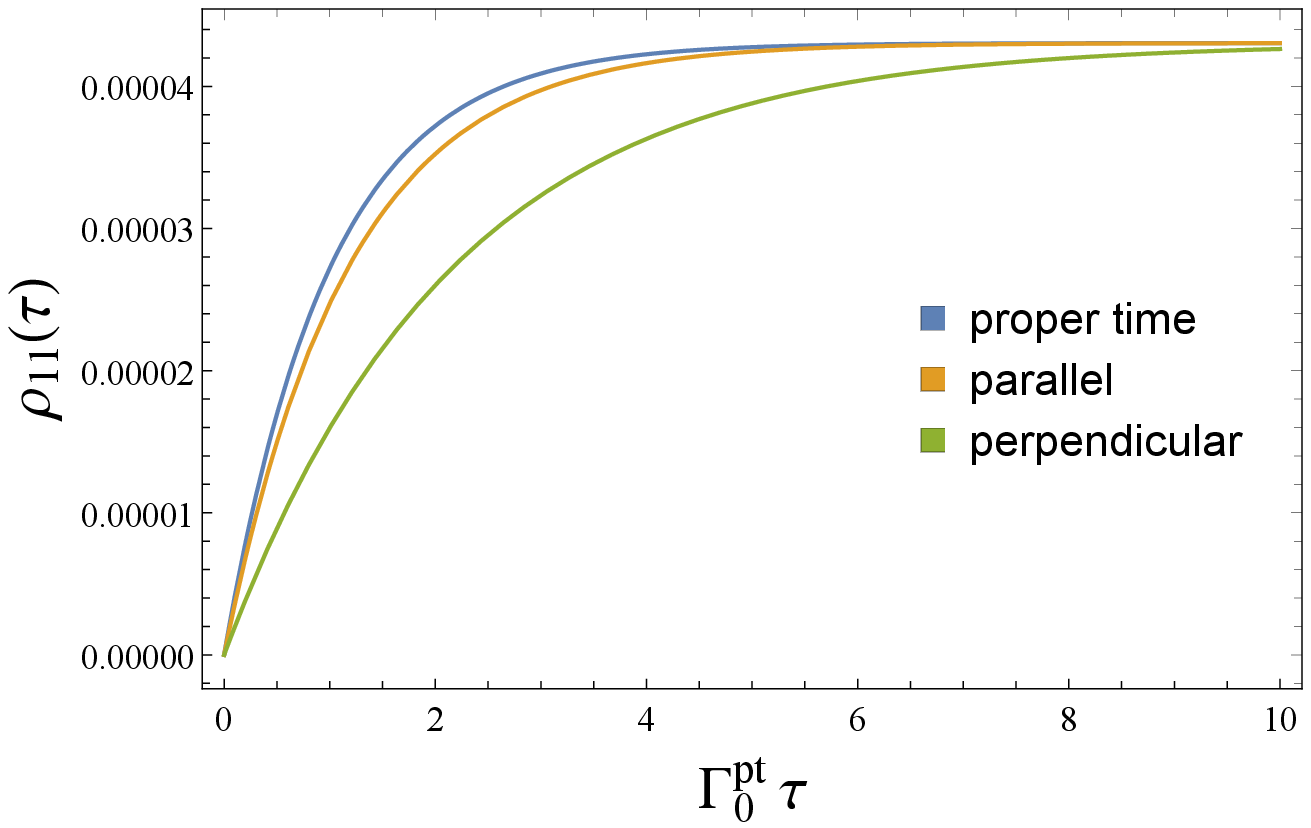}} 
\caption{Time evolution of the $\grr_{11}(\grt)$ elements (\ref{pt11}), (\ref{par11}) and (\ref{perp11}) of the density matrix of the derivative coupling detector for different values of $\grw/a$. The evolution is plotted versus a dimensionless time in units of $(\Gamma_{0}^{pt})^{-1}$, where $\Gamma_{0}^{pt}$ is the
spontaneous emission rate of the detector coupled to the proper time derivative of a scalar field at zero temperature. The detector is initially found in its ground state $\grr_{00}(0)=1$.}
\label{evol}
\end{figure}

In Fig. \ref{evol} we illustrate how the density matrix approaches its equilibrium value for the different alternatives of the derivative coupling  detector and for different values of $\grw/a$.

%=========================================
%=========================================
%=========================================
\section{Detector coupled to an electromagnetic field}\label{EMf}

We consider a 2LS detector coupled to a quantized electromagnetic field \cite{BL, Boyer, Passante, ZhuYu06}. The interaction Hamiltonian, given in the dipole approximation, reads \cite{scully, Breuer}
\be
\hat{H}_{\text{int}}=-\hat{\bf{D}}\cdot\hat{{\bf E}},
\ee
where $\hat{\bf{D}}$ is the dipole operator and $\hat{\bf{E}}$ is the electric field operator in the Schr\"{o}dinger picture
\be
\hat{{\bf E}}=i\sum_{\bf k}\sum_{\grl=1,2}\sqrt{\frac{2\pi \grw_{\bf k}}{V}}{\bf e}_{\grl}({\bf k})\left[\hat{b}_{\grl}({\bf k})-\hat{b}_{\grl}({\bf k})^{\dagger}\right].
\ee
Here, ${\bf e}_{\grl}$ is the unit polarization vector, $V$  is the volume space and $\hat{b}_{\grl}^{\dagger}, \hat{b}_{\grl}$ are the field creation and annihilation operators. Working in the interaction picture we have
\be
\hat{H}_{\text{int}}(\grt)=-\sum_i\hat{D}_i(\grt)\hat{E}_i(\grt)=
-\sum_i d_i \hat{m}(\grt)\hat{E}_i(\grt),
\ee
where ${\bf d}$ is the transition matrix element of the dipole operator, $\hat{m}({\grt})$ is the monopole moment operator (\ref{mmoment}) and $\hat{E}_i(\grt)$ denotes an electric field component in the interaction picture. For convenience, we have assumed that ${\bf d}={\bf d}^*$.

Employing the Born-Markov approximation, the time evolution equations of detector's reduced density matrix read
\bea
\dot{\grr}_{11}(\grt) =&& \sum_{i,j}d_id_j\grr_{11}(\grt)\int_0^{\infty}dt\left[e^{i\grw t}\Delta_{ij}^{+}(t)+e^{-i\grw t}\Delta_{ij}^{-}(t)\right]
\nonumber \\
&-& \sum_{i,j}d_id_j\grr_{00}(\grt)\int_0^{\infty}dt\left[e^{-i\grw t}\Delta_{ij}^{+}(t)+e^{i\grw t}\Delta_{ij}^{-}(t)\right], \label{dmme1}
\eea
\bea
\dot{\grr}_{00}(\grt) =&&\sum_{i,j}d_id_j\grr_{00}(\grt)\int_0^{\infty}dte^{-i\grw t}\left[\Delta_{ij}^{+}(t)+e^{i\grw t}\Delta_{ij}^{-}(t)\right] \nonumber \\
&-&\sum_{i,j}d_id_j\grr_{11}(\grt)\int_0^{\infty}dt\left[e^{i\grw t}\Delta_{ij}^{+}(t)+e^{-i\grw t}\Delta_{ij}^{-}(t)\right],  \label{dmme3}
\eea
\bea
\dot{\grr}_{10}(\grt)=&&\sum_{i,j}d_id_j\grr_{10}(\grt)\int_0^{\infty}dt\left[e^{i\grw t}\Delta_{ij}^{+}(t)+e^{i\grw t}\Delta_{ij}^{-}(t)\right]\nn\\&-&\sum_{i,j}d_id_j\grr_{01}(\grt)e^{2i\grw\grt}\int_0^{\infty}dt\left[e^{-i\grw t}\Delta_{ij}^{-}(t)+e^{-i\grw t}\Delta_{ij}^{+}(t)\right] \label{dmme2},
\eea
where $\Delta_{ij}^+(\grt -\grt')=\langle 0|\hat{E}_i(\grt)\hat{E}_j(\grt')|0\rangle$ or $\Delta_{ij}^+(\grt -\grt')=\langle 0|\hat{B}_i(\grt)\hat{B}_j(\grt')|0\rangle$  and $\Delta_{ij}^-(\grt -\grt')=\langle 0|\hat{E}_i(\grt')\hat{E}_j(\grt)|0\rangle$ or $\Delta_{ij}^-(\grt -\grt')=\langle 0|\hat{B}_i(\grt')\hat{B}_j(\grt)|0\rangle$ are the correlation functions of the electric or the magnetic field  components.

For a uniformly accelerated detector following the hyperbolic trajectory (\ref{atraj}) the correlation functions  read \cite{Takagi}
\be\label{corrE}
\langle 0|\hat{E}_i(\grt)\hat{E}_j(\grt')|0\rangle=\langle 0|\hat{B}_i(\grt)\hat{B}_j(\grt')|0\rangle =\frac{ a^4\grd_{ij}}{16\pi^2}
\frac{1}{\sinh^4\left[a(\grt-\grt'-i\gre)/2\right]}.
\ee
The spatial components of the electric and magnetic field operators are those parallel to the direction of motion of the detector. The same applies to the transverse direction to the direction of motion. 

The derivative coupling detector model has some features in common with the model of a detector coupled to an EM field. The correlation functions (\ref{corrE}) are similar to the correlation functions (\ref{wperp}) obtained for a UDW detector coupled to the derivative of a spatial component perpendicular to the direction of acceleration. The solutions of the evolution equations of the detector's reduced density matrix are
\begin{eqnarray}
\grr_{11}(\grt)&=&\frac{1}{2}\left[1-\frac{\Gamma^{\text{em}}_0}{\Gamma^{\text{em}}}\left(1+\frac{a^2}{\grw^2}\right) \right]+\frac{\Gamma^{\text{em}}_0}{2\Gamma^{\text{em}}}\left(1+\frac{a^2}{\grw^2}\right) e^{-\Gamma^{\text{em}} \grt}-\frac{1}{2}e^{-\Gamma^{\text{em}} \grt}(\grr_{00}(0)-\grr_{11}(0)), \label{emr11} \\
\grr_{00}(\grt) &=& 1 - \grr_{11}(\grt), \label{emr00}\\
\rho_{10}(\grt)&=&e^{-i\bar{\grw}\grt-\frac{\Gamma^{\text{em}}}{2}\grt}\grr_{10}(0)+\left[\frac{\Gamma^{\text{em}}}{2}-i(\bar{\grw}-\grw)\right]\frac{\sin(\bar{\grw}\grt)}{\bar{\grw}}e^{-\frac{\Gamma^{\text{em}}}{2}\grt}\grr_{01}(0), \label{emr10}
\end{eqnarray}
where the thermal decay constant reads
\begin{eqnarray}
 \Gamma^{\text{em}} =  \Gamma^{\text{em}}_0 \left(1+\frac{a^2}{\grw^2}\right)  \coth\left(\frac{\pi\grw}{a}\right) \label{decaya}
\end{eqnarray}
and the decay constant in the static case is 
\be
\Gamma^{\text{em}}_0 = \frac{|{\bf d}|^2\grw^3}{3\pi}.
\ee
The shifted frequency  is
\be
\bar{\grw}=\grw+\frac{\Gamma^{\text{em}}_0}{\pi}\left(1+\frac{a^2}{\grw^2}\right)\left(\log(e^{\grg}\gre\grw)+\log(a/\grw )+Re\left\{\grc\left(\frac{i\grw}{a}\right)\right\}\right)+\frac{\Gamma^{\text{em}}_0}{\pi}C'_R.
\ee

Evaluating the early-time transition rate we obtain
\be\label{emtrans}
w_{em}=\Gamma^{\text{em}}_0 \left(1+\frac{a^2}{\grw^2}\right)\frac{1}{e^{2\pi\grw/a}-1},
\ee
which is in a non-Planckian form. The transition rate (\ref{emtrans}) is identical to that of a static 2LS interacting with an EM field that is in thermal equilibrium at $T_U$ \cite{Carmichael}. In the long-time limit ($\Gamma t >> 1$), the density matrix of the detector approaches the equilibrium value
\be
\hat{\rho}_{\infty} = \left( \begin{array}{cc} \frac{e^{-\frac{2\pi\grw}{a}}}{e^{-\frac{2\pi\grw}{a}}+1}& 0 \\0 &\frac{1}{e^{-\frac{2\pi\grw}{a}}+1} \end{array} \right), 
\ee
even in the non-Markovian regime.

%=========================================
%=========================================
%=========================================
\section{Discussion and conclusions}\label{concl}

\begin{table}[t!]
\begin{center}
\begin{tabular}{|c||c|c||c|}
\hline
Accelerating detector coupled to &  Transition rate  & Planckian form &Thermal asymptotic state \\
\hline \hline
scalar field & $\frac{g^2\grw}{2\pi}\frac{1}{e^{\frac{2\pi\grw}{a}}-1}$  & $ \checkmark$ & \checkmark \\
\hline
 proper time derivative & $\frac{g^2\grw^3}{2\pi}\frac{1}{e^{\frac{2\pi\grw}{a}}-1}$ & $\checkmark$ & \checkmark\\
\hline
direction parallel to acceleration & $\frac{g^2\grw^3}{6\pi}\left(1+\frac{4a^2}{\grw^2}\right)\frac{1}{e^{\frac{2\pi\grw}{a}}-1}$  & $ \times$ & \checkmark \\
\hline
 direction perpendicular to acceleration &  $\frac{g^2\grw^3}{6\pi}\left(1+\frac{a^2}{\grw^2}\right)\frac{1}{e^{\frac{2\pi\grw}{a}}-1}$ &$\times$ & \checkmark\\
\hline
EM field& $\frac{g^2\grw^3}{3\pi} \left(1+\frac{a^2}{\grw^2}\right)\frac{1}{e^{\frac{2\pi\grw}{a}}-1}$ & $\times$ & \checkmark \\
\hline
\end{tabular}
\end{center}
\caption{Comparison between transition rate and asymptotic state of accelerated detectors for different types of interaction between the detector and the field,  within the Markovian regime of high accelerations. When non-Markovian effects are taken into account the transition rate is non-Planckian in all situations.} \label{results}
\end{table}

Employing the notion of an Unruh-DeWitt detector, the Unruh effect can be expressed entirely in terms of {\em local} physics. The transition rate of a uniformly accelerated  detector is evaluated to leading order in perturbation theory and is found to obey a Planck distribution at the Unruh temperature. This feature of the transition rate is usually considered as validation of the Unruh effect.

However, the perturbative evaluation of the transition rate has a restricted domain of applicability. For quantum field probes, such as the Unruh-DeWitt detectors, it applies {\em only} during very early times. This is because it ignores the effect of spontaneous emission after excitations. Perturbative evaluation also ignores the backaction of the field to the detector. In order to take these effects into account, the Unruh-DeWitt detectors should be treated as {\em open quantum systems}.

Open quantum systems are generally described by a second order master equation, which is derived implementing the
Born-Markov approximation and RWA. The second-order master equation is an excellent approximation to a large class of problems, but it turns out that the Markov approximation has some limitations. For example, the Markovian master equation is not valid for low temperatures of the environment \cite{vega}. Indeed, when studying the response of moving detectors interacting with quantum fields, non-Markovian effects are particularly pronounced  at early times and for small accelerations \cite{DMCA,Lin}. Thus, the Markov approximation cannot be presupposed as in \cite{{Benatti}}, and is applied only for high accelerations.

In this paper, we treated an Unruh-DeWitt detector as an open quantum system, with a quantum field playing the role of the environment, and evaluated the response of a uniformly accelerated detector for different types of interaction between the detector and the field. We found that the early-time transition rate strongly depends on the type of the interaction and may not be Planckian, even in the Markovian regime of high accelerations. In  contrast, the asymptotic state of an accelerated detector is always thermal at the Unruh temperature, regardless the internal characteristics of the interaction or the interacting field. The detector's density matrix at late times is thermal even if we take into account non-Markovian effects. Our  results are summarized in Table \ref{results}.

Our work strongly implies that the asymptotic state of an Unruh-DeWitt detector provides a more fundamental and persistent characterization of the acceleration temperature: A uniformly accelerated detector experiences the field vacuum as a {\em genuine thermal bath} at the Unruh temperature $T_U = a/(2\pi)$ and eventually settles at a thermal state, regardless their intermediate dynamics or the type of interaction.

%=====================================================================================================
% ACKNOWLEDGMENTS
%=====================================================================================================
\begin{acknowledgments}

The author wishes to thank Charis Anastopoulos, Benito A. Ju\'arez-Aubry and Jorma Louko for helpful comments and discussions. This research has been cofinanced--via a programme of State Scholarships Foundation (IKY)--by the European Union (European Social Fund-ESF) and Greek national funds through the action entitled ``Scholarships programme for postgraduates studies -2nd Study Cycle" in the framework of the Operational Programme``Human Resources Development Program, Education and Lifelong Learning" of the National Strategic Reference Framework (NSRF) 2014--2020. Early stages of the work were supported by Grant No. E611 from the
Research Committee of the University of Patras
via the ``K. Karatheodoris" program.

\end{acknowledgments}
%=============================================
%=============================================
%APPENDIX
%=============================================
%=============================================
\appendix

\section{Integral transforms of Wightman functions}\label{LPT}

In this appendix, we evaluate the Laplace transform integrals of the Wightman functions that are used in our calculations. 
First, we calculate the integral 
\bea\label{lpint}
I&=&\int_0^{\infty}e^{-z\grt}\frac{1}{\sinh^2[a(\grt-i\gre)/2]}  d\grt \nn \\
&=&-\frac{2}{a}\int_0^{\infty}e^{-z\grt}d\left\{\coth[a(\grt-i\gre)/2]\right\}\nn\\
&=&-\frac{2}{a}\left\{\coth[(ia\gre)/2]+z\int_0^{\infty}e^{-z\grt}\coth[a(\grt-i\gre)/2]d\grt\right\}.
\eea

The integral
\be\label{LaplInt}
I_1=\int_0^{\infty}e^{-z\grt}\coth[a(\grt-i\gre)/2]d\grt.
\ee
is evaluated  as
\bea
I_1&=&\int_0^{\infty}e^{-z\grt}\left[1-e^{-a(\grt-i\gre)}\right]^{-1}d\grt+\int_0^{\infty}e^{-a(\grt-i\gre)}e^{-z\grt}\left[1-e^{-a(\grt-i\gre)}\right]^{-1}d\grt\nn \\
&=&\frac{1}{a}\int_0^1t^{\frac{z}{a}-1}\left[1-e^{ia\gre}t\right]^{-1}dt+\frac{e^{ia\gre}}{a}\int_0^1t^{\frac{z}{a}}\left[1-e^{ia\gre}t\right]^{-1}dt\nn \\
&=&\frac{1}{z}\ {}_2F_1\left(1,\frac{z}{a},\frac{z}{a}+1;e^{ia\gre}\right)+\frac{e^{ia\gre}}{z+a}\ {}_2F_1\left(1,\frac{z}{a}+1,\frac{z}{a}+2;e^{ia\gre}\right),
\eea
where ${}_2F_1(a,b,c;w)$ is the Gauss hypergeometric function \cite{Abramowitz,Ryzhik}. We used the integral representation of the hypergeometric function
\be
{}_2F_1(a,b,c;w)=\frac{\Gamma(c)}{\Gamma(b)\Gamma(c-b)}\int_0^1 t^{b-1}(1-t)^{c-b-1}(1-tw)^{-a}dt.
\ee
Thus, the integral (\ref{lpint}) is
\bea\label{LTpos}
I=-\frac{2}{a}\Big\{&&\coth[(i\gra\gre)/2]+{}_2F_1\left(1,\frac{z}{\gra},\frac{z}{\gra}+1;e^{i\gra\gre}\right)\nn\\&&+\frac{ze^{ia\gre}}{z+a}\ {}_2F_1\left(1,\frac{z}{\gra}+1,\frac{z}{\gra}+2;e^{i\gra\gre}\right)\Big\}.
\eea

The hypergeomertic series ${}_2F_1(a,b,c;w)$ is analytic everywhere in the complex plane except for the branch points at $w=0,1,\infty$. When $w\to1$, the zero-balanced hypergeometric series, i.e., the series that $c-b-a=0$, behave as
\be
\frac{\Gamma(a)\Gamma(b)}{\Gamma(a+b)}\ {}_2F_1(a,b,a+b;w)=-2\grg-\psi(a)-\psi(b)-\log(1-w)+o(1),
\ee
where $\grg$ is the Euler-Mascheroni constant and $\psi(z)=\frac{d}{dz}\log\Gamma(z)$ is the digamma (psi) function \cite{Abramowitz}. Expanding around the branch point $w=1$, we write the integral (\ref{LTpos}) as
\be
I=\mathcal{L}\left\{\sinh^{-2}[a(\grt-i\gre)/2]\right\}(z)=-\frac{2}{a}\left\{\coth\left(\frac{ia\gre}{2}\right)-1-\frac{2z}{a}\left[\log(e^{\grg}\gre a)+\grc\left(\frac{z}{a}\right)\right]+\frac{i\pi z}{a}\right\}
\ee
where we used $\psi(1)=-\gamma$ and $\psi(1+z)=1/z+\psi(z)$. The logarithm is taking values in the principal branch. We also have
\be
I=\mathcal{L}\left\{\sinh^{-2}[a(\grt+i\gre)/2]\right\}(z)=-\frac{2}{a}\left\{-\coth\left(\frac{ia\gre}{2}\right)-1-\frac{2z}{a}\left[\log(e^{\grg}\gre a)+\grc\left(\frac{z}{a}\right)\right]-\frac{i\pi z}{a}\right\}
\ee

The correlation function in the case of UDW detector coupled to the proper time derivative of a scalar field reads
\be
W^{\pm}(\grt-\grt')=\partial_{\grt}\partial_{\grt'}\frac{1}{\sinh^{2}[a(\grt-\grt'\mp i\gre)/2]}=-\frac{d^2}{dt^2}\frac{1}{\sinh^{2}[a(t\mp i\gre)/2]}
\ee
Taking the Laplace transform we have
\begin{align}\label{LPTdr}
\mathcal{L}\left\{W^+(t)\right\}(z)=\frac{2z^2}{a}\bigg\{&\coth\left(\frac{ia\gre}{2}\right)-1-\frac{2z}{a}\left[\log(e^{\grg}\gre a)+\grc\left(\frac{z}{a}\right)\right]+\frac{i\pi z}{a}\nn\\
\ &+\frac{z}{\sinh^2(i\gre a/2)}-\frac{a\cosh(i\gre a/2)}{\sinh^3(i\gre a/2)}\bigg\},
\end{align}
where we have used the Laplace transform identity $\mathcal{L}\{f''(t)\}(z)=z^2\mathcal{L}\{f\}(z)-zf(0)-f'(0)$. Similarly, we have
\begin{align}\label{ltb}
\mathcal{L}\left\{W^-(t)\right\}(z)=\frac{2z^2}{a}\bigg\{&-\coth\left(\frac{ia\gre}{2}\right)-1-\frac{2z}{a}\left[\log(e^{\grg}\gre a)+\grc\left(\frac{z}{a}\right)\right]-\frac{i\pi z}{a}\nn\\
\ &+\frac{z}{\sinh^2(i\gre a/2)}+\frac{a\cosh(i\gre a/2)}{\sinh^3(i\gre a/2)}\bigg\}.
\end{align}

%=========================================
%=========================================
%=========================================
%BIBLIOGRAPHY
%=============================================
%=============================================

\end{document}